\documentclass{ptephy_v1}

\usepackage{latexsym,graphicx,amssymb,amsmath,mathrsfs}
\usepackage{epstopdf}
\usepackage{url}

\graphicspath{{./fig/}}

\newcommand\APF{\langle{e^{i\theta}}\rangle_\mathrm{pq}}

\begin{document}

\preprintnumber{KUNS-2704, YITP-17-98}

\title{
Application of neural network to sign problem via
path optimization method
}

\author[1]{Yuto Mori}
\affil{Department of Physics, Faculty of Science, Kyoto
University, Kyoto 606-8502, Japan\email{mori.yuto.47z@st.kyoto-u.ac.jp}}

\author[2]{Kouji Kashiwa}
\affil{Yukawa Institute for Theoretical Physics,
Kyoto University, Kyoto 606-8502, Japan
\email{kouji.kashiwa@yukawa.kyoto-u.ac.jp}
\email{ohnishi@yukawa.kyoto-u.ac.jp}}

\author[2]{Akira Ohnishi}

\begin{abstract}
 We introduce the feedforward neural network to attack
 the sign problem via the path optimization method.
 The variables of integration is complexified and the integration
 path is optimized in the complexified space by minimizing the cost
 function which reflects the seriousness of the sign problem.
 For the preparation and optimization of the integral path in
 multi-dimensional systems, we utilize the feedforward neural network.
 We examine the validity and usefulness of the method in
 the two-dimensional complex $\lambda \phi^4$ theory
 at finite chemical potential as an example of the quantum field theory
 having the sign problem.
 We show that the average phase factor is
 significantly enhanced after
 the optimization and then we can safely perform the hybrid Monte-Carlo
 method.
\end{abstract}

\maketitle

\section{Introduction}

The feedforward neural
network~\cite{mcculloch1943logical,hebb2005organization,
rosenblatt1958perceptron,hinton2006reducing}
is a widely used mathematical model in the machine learning.
Its name is originated from similarities with animal brains;
there are the input, hidden and output layers and their connections
replicate brains.
In the model, we have some parameters and these are used to learn
training data and/or minimize the function which is so called
the cost function.
Therefore, this model can be used in the optimization problems
and thus we can apply it to attack the sign problem.

The sign problem is 
caused by the integral of rapidly oscillating function,
and thus it appears in
many fields in physics
such as the elementary particle physics, hadron and nuclear physics
and the condensed matter physics.
Particularly, the sign problem has attracted much more
attention recently in Quantum Chromodynamics (QCD) at finite baryon
density because there are remarkable progress based on the complex
Langevin method~\cite{Parisi:1980ys,Parisi:1984cs} and the Lefschetz
thimble method~\cite{Witten:2010cx,Cristoforetti:2012su,Fujii:2013srak,Alexandru:2015sua,Alexandru:2017czx}.
In the Complex Langevin method, we generate configurations by solving
the complex Langevin equation, and the sign problem does not appear.
However, the Complex Langevin method may provide wrong results
when the drift term shows singular behavior~\cite{Aarts:2009uq,Nishimura:2015pba}.
In comparison, the Lefschetz-thimble method is based on the
Picard-Lefschetz theory~\cite{Witten:2010cx} and within the standard
path-integral formulation.
We construct the integral path referred to as the Lefschetz thimbles
by solving the holomorphic flow starting from fixed points.
Then the partition function can be expressed by
the sum of contributions from relevant Lefschetz thimbles.
Relevance of a thimble can be determined by the crossing behavior
of the dual thimble, the holomorphic flow to the fixed point.
Relevant thimbles can be also obtained by modifying the
integral path from the original integral path according to the holomorphic flow~\cite{Alexandru:2015sua}.
On each Lefschetz thimble, the imaginary part of the action is
constant, but the Jacobian generated by the bending
structure of the integral path causes a phase
(residual sign problem).
In addition, there is the cancellation between
different relevant thimbles (global sign problem).
Recently, one more serious problem in the Lefschetz-thimble method
has been discussed~\cite{Mori:2017zyl}.
If the action has the square root or the logarithm,
there appear singular points and cuts on the complexified variables
of integration, which obstruct to draw continuous Lefschetz-thimbles
in the numerical calculation of holomorphic flows.
Thus it is valuable to develop a method, in which an appropriate integral path is obtained without suffering from the singular points of the action.

Recently, the present authors have proposed a new method
to attack the sign problem which is so called
the path optimization method~\cite{Mori:2017pne}.
The path optimization method is based on the complexification of
variables of integration.
We prepare the trial function representing the modified integral path, 
which is given as a simple function having a few parameters.
The parameters are tuned to minimize the cost function which reflects
the seriousness of the sign problem.
In Ref.~\cite{Mori:2017pne}, we have demonstrated that the optimized path matches the Lefschetz thimbles
around the fixed points in a one-dimensional model with a serious sign problem.
In systems having large degrees of freedom,
it is too tedious to prepare
the functional form of the
trial function by hand,
then it is helpful to apply the machine learning technique
to obtain the optimized integral path.

Machine learning has been recently utilized to attack
the sign problem~\cite{Ohnishi:2017zxh,Alexandru:2017czx}.
In Ref.~\cite{Ohnishi:2017zxh}, the mono-layer neural network is adopted,
and the optimized path
is found to agree with that in Ref.~\cite{Mori:2017pne}.
In Ref.~\cite{Alexandru:2017czx}, the multi-layer neural network is applied to
Thirring model with supervision by some of the configurations from
generalized Lefschetz thimble method.

In this article,
we use the feedforward neural network with a hidden mono-layer
to apply the path optimization method to a quantum field theory
with a sign problem, 
the complex $\lambda \phi^4$ theory with finite chemical potential.
The average phase factor, the scatter plot of fields and the expectation
value of the number density are investigated.
Some attempts have been done to apply the machine learning to
QCD~\cite{Csabai:1991ee,Forte:2002fg,DelDebbio:2007ee},
but our application is a bit different;
our machine learning is the unsupervised learning and thus the process
is similar to that for board
games~\cite{tesauro1992practical,silver2016mastering} rather
than that for image recognitions~\cite{krizhevsky2012imagenet};
we optimize our integral path in the plane of complexified variables of
integration to aim to {\it win the sign problem}.

This paper is organized as follows.
In Sec.~\ref{Sec:POM},
we explain the formalism of the path optimization method.
Section~\ref{Sec:NN} shows details of the feedforward neural network.
In Sec.~\ref{Sec:phi4}, we examine the method in two-dimensional
$\lambda\phi^4$ theory.
The action and the numerical setup are
explained in Sec.~\ref{Sec:action}
and Sec.~\ref{Sec:Numerical_setup}, respectively, and
results are shown in Sec.~\ref{Sec:Numerical_results}.
Section~\ref{Sec:Summary} is devoted to summary.

\section{Path optimization method}
\label{Sec:POM}

The path optimization method is the method of varying the integral path
in the plane of complexified variables of integration to minimize the
cost function~\cite{Mori:2017pne}; the cost function should reflect
the seriousness of the sign problem and then the optimization process
controls how much the numerical integration is precise.
Actual procedure is given as follows.
For the theory with $n$ degree of freedoms, we start form
the complexification of variables of
integration with parametric variables $t$,
$x_i \in \mathbb{R} \rightarrow z_i(t) \in \mathbb{C}$ where
$i = 1, \cdots n$.
Next, we set the trial function containing some parameters
to represent the modified integral path.
The optimization of the cost function is then performed
by tuning parameters in the trial function.

One of the candidates for the cost function is given as
\begin{align}
 {\cal F}[z(t)]
 &= \frac{1}{2} \int d^nt~ |e^{i \theta(t)}
   - e^{i \theta _0}|^2 \times |J(t) e^{-S(z(t))}| \nonumber \\
  &= \int d^nt~|J(t) e^{-S(z(t))}|
   - \left |\int d^nt~J(t) e^{-S(z(t))} \right |
\nonumber\\
 &=\left| \mathcal{Z} \right|
 \left[ |\langle e^{i\theta}\rangle_\mathrm{pq}|^{-1} - 1\right]
 \ ,
\label{Eq:cf}
\end{align}
where ${\cal Z}$ is the partition function,
$J(t) = \det(\partial z_i/\partial t_j)$ is the Jacobian,
and $\langle e^{i\theta} \rangle_\mathrm{pq}$ means
the average phase factor,
\begin{align}
 \mathcal{Z}&= \int d^nt\, J(t) e^{-S(z(t))}\ ,\\
 \theta(t) &= \arg (J(t)e^{-S(z(t))}), ~~~~
 \theta_0   = \arg ({\cal Z}), \\
 \langle \mathcal{O} \rangle _\mathrm{pq} &=
  \frac{\int d^nt~\mathcal{O}(z(t))|J(t)e^{-S(z(t))}|}
       {\int d^nt~|J(t)e^{-S(z(t))}|}\ .
\end{align}
The subscripts $\mathrm{pq}$ means the phase quenched average.
The first term in the cost function, Eq.~(\ref{Eq:cf}),
is proportional to the inverse average phase factor
$\langle e^{i\theta} \rangle_{pq}^{-1}$,
Then this cost function represent the seriousness of the sign problem
and depends on the detailed shape of the deformed integral path.

In calculation of the cost function, Eq.~(\ref{Eq:cf}), 
we assume that integral of analytic function is independent of the path
due to Cauchy's integral theorem.
Integrals on two different paths are the same
as long as the contribution from infinity $|z| = \infty$ is zero
and the deformed path does not go across the singular points.
In this paper, we treat the complex $\phi^4$ theory, and the action decreases
more rapidly than gaussian at
$|\mathrm{Re}~z| \rightarrow \infty, |\mathrm{Im}~z| < \infty$.
Thus, we restrict the integral path so that imaginary parts of variables
are finite.
Under this condition, expectation values of observables,
\begin{align}
\langle \mathcal{O} \rangle = 
 \frac{\langle \mathcal{O}e^{i\theta} \rangle _{pq}}
      {\langle e^{i\theta} \rangle _{pq}},
\end{align}
are also path-independent.

After setting the cost function,
the next task is arrangement of the functional form for
the trial function.
In Ref.~\cite{Mori:2017pne}, a simple trial function is adopted
and the optimization is demonstrated to be successful in
a one variable model.
By comparison,
it is natural to think that the task becomes difficult
in complicated systems having large degrees of freedom such as
the field theory.
To circumvent the problem, we use the feedforward neural network
as a possible and practical solution
instead of arrangement of the functional form for the trial function.

\section{Neural network}
\label{Sec:NN}

\begin{figure}[htb]
\begin{center}
 \includegraphics[width=0.4\textwidth]{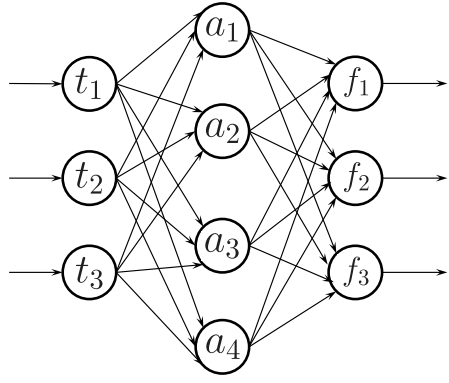}
\end{center}
 \caption{
 Schematic figure of the feedforward neural network.
 Arrows indicates the propagation of signals.
 }
\label{Fig:FNN}
\end{figure}

In this section, we give a brief review on the feedforward neural
network and explain how to use it for the path optimization method.
The feedforward neural network is the basic algorithm
containing the input, hidden and output layers as shown in
Fig.~\ref{Fig:FNN}.
Each layer has many units, and units of two layers are bonded as
represented by arrows in the figure.
Input values are $t_i$, output values are $f_i$, and values of the hidden layer is $a_i$.
Then, values of the hidden and output layers are given by
\begin{align}
a_i = g(W^{1}_{ij}t_{j} + b^{1}_i), \nonumber \\
f_i = g(W^{2}_{ij}a_{j} + b^{2}_i),
\end{align}
where $W$ and $b$ are parameters for the
optimization and characterize the feedforward neural network.
The function $g(x)$ is non-linear function
which is so called the activation function;
the sigmoid function, $g(x) = 1/(1+e^{-x})$, and hyperbolic
tangent are typical examples.

It should be noted that we can use a {\it deep} neural network
having three or more hidden layers,
but the hidden mono-layer is found to be enough
to perform the optimization in the present computation:
The feedforward neural network even with the hidden mono-layer can
simulate
any kind of continuous functions on the compact subset
as long as 
we can prepare sufficient number of units in the
hidden layer, as given in the universal approximation
theorem~\cite{cybenko1989approximation,hornik1991approximation}.

In practice,
we regard $t$ ($f$) as an input (output) and give the integral path as
\begin{align}
z_i(t) = t_i + i (\alpha_i f_i(t) + \beta_i)
\ ,
\label{Eq:tri}
\end{align}
where $\alpha_i$ and $\beta_i$ are parameters. 
For simplicity, we assume that the real part is not modified.
We use the hyperbolic tangent
for the activation function in the feedforward neural network.
In the updation of $\alpha_i, \beta_i$ and parameters in $f_i$, we employ the back-propagation
algorithm~\cite{rumelhart1988learning}.

To evaluate the cost function by using the feedforward neural network
with the hybrid Monte-Carlo method,
we normalize Eq.~(\ref{Eq:cf}) by $\int d^nt~P_0(t)$ with an
appropriate probability distribution $P_0(t)$,
\begin{align}
  \frac{{\cal F}[z(t)]}{\int d^n t ~ P_0(t)} =
  & \frac{\int d^n t \frac{|Je^{-S}|}{P_0}P_0}{\int d^n t ~ P_0}
   - \left| \frac{\int d^n t \frac{Je^{-S}}{P_0}P_0}{\int d^n t ~ P_0} \right| \nonumber \\
  \simeq & \frac{1}{N} \sum_{k}^{N} \frac{|J(t^{(k)})e^{-S(z(t^{(k)}))}|}{P_0(t^{(k)})}
  - \frac{1}{N} \left| \sum_{k}^{N} \frac{J(t^{(k)})e^{-S(z(t^{(k)}))}}{P_0(t^{(k)})} \right| , \label{eq:cf2}
\end{align}
where $\{t^{(k)}\}$ $k$-th configuration
sampled according to $P_0(t)$,
and $N$ is the number of configurations.
Let $c=\{c_i\}$ be the set of parameters appearing in the feedforward neural network,
$z(t) = z(t, c) $,
then derivatives of ${\cal F}$ with respect to $c_i$ are evaluated as
\begin{align}
  \frac{\partial}{\partial c_i} \frac{{\cal F}[z(t)]}{\int d^n t ~ P_0(t)} &\simeq
   \frac{1}{N} \sum_{k}^{N} F_i(t^{(k)}, c), \label{eq:diff}
\end{align}
where the derivative for each configuration, $F_i$, is given as
\begin{align}
F_i(t, c) &=
   \frac{|J(t)e^{-S(z(t))}|}{P_0(t)}~
   {\rm Re} \left[ (1-e^{i(\theta(t) - \theta_0)})
   \frac{\partial}{\partial c_i} \log(J(t)e^{-S(z(t))}) \right] . \label{eq:diff2}
\end{align}
Here we have used the relation
$\arg(\sum_k J(t^{(k)})e^{-S(z(t^{(k)}))}/P_0(t^{(k)}))
\simeq \arg({\cal Z}) = \theta_0$, which holds for a large number of
Monte-Carlo configurations.

Because Eq.~(\ref{eq:diff}) is expressed as a summation,
the stochastic gradient descent (SGD) is available for the optimization.
In this study, we employ the ADADELTA algorithm~\cite{zeiler2012adadelta}
which is the extended algorithm of SGD
with preventing the learning weight decay as the optimizer,
and it is compared
with the Adam algorithm~\cite{kingma2014adam} to check the stability.
In ADADELTA algorithm, we update the parameters in the fictitious time step $j$, $c_i^{(j)}$, as
\begin{align}
c_i^{(j+1)} = c_i^{(j)} - \eta v_i^{(j+1)},
\end{align}
where $\eta$ is a parameter so called the learning rate.
The direction of the update, $v_i$, is obtained as,
\begin{align}
v_i^{(j+1)} &= \frac{\sqrt{s_i^{(j)}+\epsilon}}{\sqrt{r_i^{(j+1)}+\epsilon}} F_i^{(j)}, \\
r_i^{(j+1)} &= \gamma r_i^{(j)} + (1-\gamma) (F_i^{(j)})^2, \\
s_i^{(j+1)} &= \gamma s_i^{(j)} + (1-\gamma) (v_i^{(j+1)})^2,
\end{align}
where
$\gamma$ is the decay constant, and $\epsilon$ is a positive small number
introduced to avoid the divergence.
The variables $r$ and $s$ denote the mean square values
of $F_i$ and $v_i$, respectively,
and we set $r^{(0)}=0,~s^{(0)}=0$ for the initial condition.
In the ($j+1$)-th update, we evaluate the variables in the order
of $r_i^{(j+1)}$, $v_i^{(j+1)}$ and $s_i^{(j+1)}$.
After updates of many times,
we find
$v_i/\sqrt{\langle{v_i^2+\epsilon}\rangle}\simeq F_i/\sqrt{\langle{F_i^2+\epsilon}\rangle}$,
where $\langle\cdots\rangle$ shows the average over the updates.
The optimization algorithm shown here has been found to be successful empirically,
and should be used with care.
Nevertheless, parameters converge when all the derivatives $F_i$ become zero,
as in the standard gradient methods.

In the actual optimization process, we use mini-batch training to
make our optimization faster and easier:
We divide the configurations as $N = K N_\mathrm{batch}$
where $N_\mathrm{batch}$ is the batch size and the learning is performed
in batch by batch.
To include all updation of each batch,
the parameters in the feedforward neural network is then updated by using
the mean value of $F_i$ as
\begin{align}
F_i \to
\frac{1}{N_\mathrm{batch}} \sum _{k=1}^{N_\mathrm{batch}} F_i(t^{(k)}, c).
\end{align}
In one optimization procedure we perform updations $K$ times
with $N_{\mathrm{batch}}$ configurations.

\section{Application to two-dimensional complex $\lambda \phi^4$ theory}
\label{Sec:phi4}

\subsection{Two-dimensional Complex $\lambda \phi^4$ theory}
\label{Sec:action}

In order to examine the applicability and usefulness of 
the path optimization method in quantum field theories,
we consider here the two-dimensional complex $\lambda \phi^4$ theory
with finite chemical potential
($\mu$)~\cite{Aarts:2008wh,Gattringer:2012df,Cristoforetti:2013wha,Fujii:2013sra}
as an example.
The action is defined in the lattice unit as
\begin{align}
S = \sum_x & \left [ \frac{(4+m^2)}{2}\phi_{a,x}\phi_{a,x}
 + \frac{\lambda}{4}(\phi_{a,x}\phi_{a,x})^2
 - \phi_{a,x}\phi_{a,x+\hat{1}} \right. \nonumber \\
& \left. - \phi_{a,x}\phi_{b,x+\hat{0}}(\delta_{ab}\cosh(\mu)-i\epsilon_{ab}\sinh(\mu))
 \right]
\label{Eq:action}
\end{align}
where
$a,b=1,2$, and
$\phi_1,~\phi_2 \in \mathbb{R}$ are field variables.
we assume these variables satisfy the periodic boundary condition.
In the path optimization method, we complexify variables, 
$\phi_{a,x} \to z_{a,x}\in \mathbb{C}$, with Eq.~(\ref{Eq:tri}).
Then the number density is given by 
\begin{align}
n = \frac{1}{V} \sum_x (\delta_{ab} \sinh(\mu) -i \epsilon_{ab}\cosh(\mu))z_{a,x}z_{b,x+\hat{0}},
\end{align}
where $V$ is the lattice volume, $V=L_tL_x$.

It should be noted that the action Eq.~\eqref{Eq:action} has the global $\mathrm{U}(1)$ symmetry:
The action is invariant under the rotation in the $(\phi_1,\phi_2)$ plane,
$(\phi_1,\phi_2) \to (\phi_1\cos{q}-\phi_2\sin{q},\phi_1\sin{q}+\phi_2\cos{q})$,
where $q$ is a rotation angle.
Then the distribution of $(\phi_1, \phi_2)$ should be, in principle, invariant under the $\mathrm{U}(1)$ transformation even after complexification.
In one set of Monte-Carlo samples of configurations, however, 
the distribution can break the $\mathrm{U}(1)$ symmetry spontaneously.
We will discuss this point in Subsec.~\ref{Sec:Numerical_results}.

\subsection{Numerical setup}
\label{Sec:Numerical_setup}

We have investigated the range of $\mu$ from $0.2$ to $2.0$ with an interval of
$\Delta \mu = 0.2$.
Parameters in the theory are fixed as $m = 1$ and $\lambda = 1$.
Under this setup, the mean field treatment predicts
the number density starts to increase at around $\mu=1$.

The lattice size used in this article are $L=L_t = L_x = 4$, $6$
and $8$.
The step size of the molecular dynamics in the hybrid Monte-Carlo method
is set to $\Delta \tau = \tau/n_\mathrm{step} = 0.1/10 = 0.01$,
where $n_{\mathrm{step}}$ is the number of steps.
In the optimization procedure, the number of units in the hidden layer,
$N_\mathrm{unit}$, is set to $2^5$, $2^6$ and $2^8$ for $L = 4$, $6$ and $8$,
respectively.
For ADADELTA algorithm,
we use $\eta=0.1/N_\mathrm{ite}$ where
$N_\mathrm{ite} = 1, \cdots$ is the number of the optimization,
$\gamma = 0.95$ and $\epsilon = 10^{-6}$.
In the mini-batch training, we use $N_\mathrm{batch}=10$.

To calculate the expectation values, we have generated $10^4$
configurations for each optimization,
and the expectation values are estimated after $N_\mathrm{ite}=20$
optimizations for $L=4$ and $6$
with discarding the first $2000$ configurations for thermalization.
In the case with $L=8$, we use $N_\mathrm{ite}=30$.
Statistical errors are obtained by using the Jack-knife method.

In actual numerical calculation to obtain the optimized integral path,
we perform the following procedures:
\begin{enumerate}
  \item Take initial parameters in
        Eq.~(\ref{Eq:tri}) as parameters obtained in smaller $\mu$ or
        randomly with the Xavier initialization~\cite{glorot2010understanding}.
  \item Generate configurations with
        the phase-quenched probability,
        $P_0(t)=|Je^{-S}|_{z=z_0(t)}$ where $z_0(t)$ is the original path.
        A sufficient number of configurations should be generated.
  \item Select configurations randomly and compute
	$\sum^{N_{batch}}F_i(t,c)$ to update parameters in
        Eq.~(\ref{Eq:tri}) by using the back-propagation with the
        mini-batch training.
  \item Regenerate configurations with
        $P_0(t)=|Je^{-S}|_{z=z(t)}$ where $z(t)$ is the modified path.
  \item Check the average phase factor.
	In the case where the average phase factor is not large enough
	and the iteration number, $N_\mathrm{ite}$, is still small,
	go back to step 3.
\end{enumerate}

\subsection{Numerical results}
\label{Sec:Numerical_results}

\begin{figure}[htbp]
\begin{center}
\includegraphics[width=0.45\textwidth]{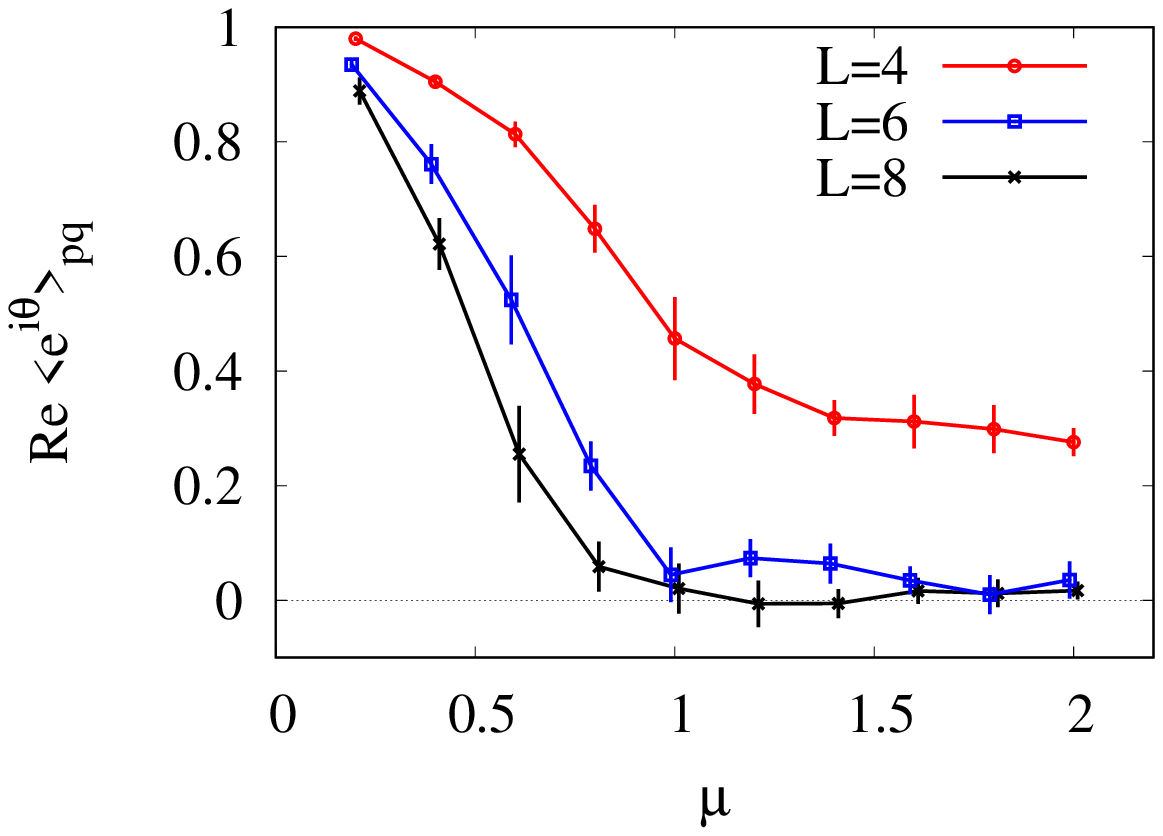}
\includegraphics[width=0.45\textwidth]{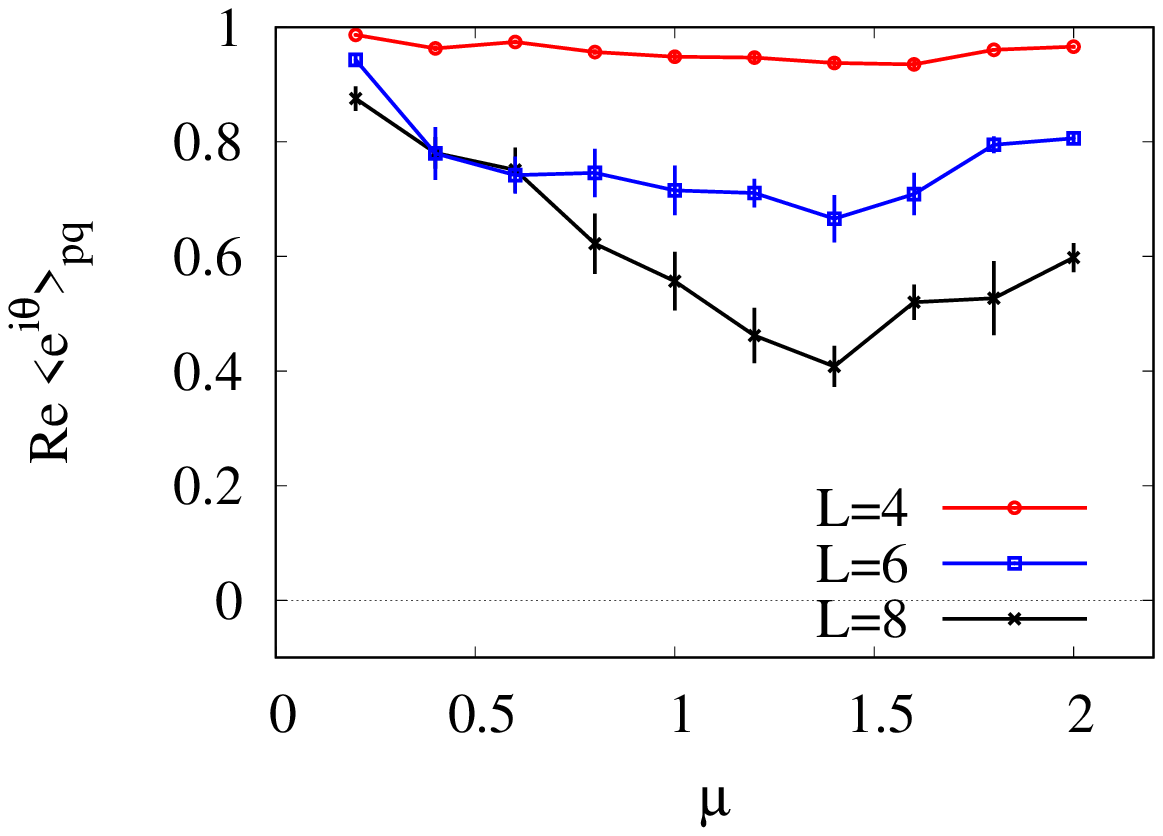}
\includegraphics[width=0.45\textwidth]{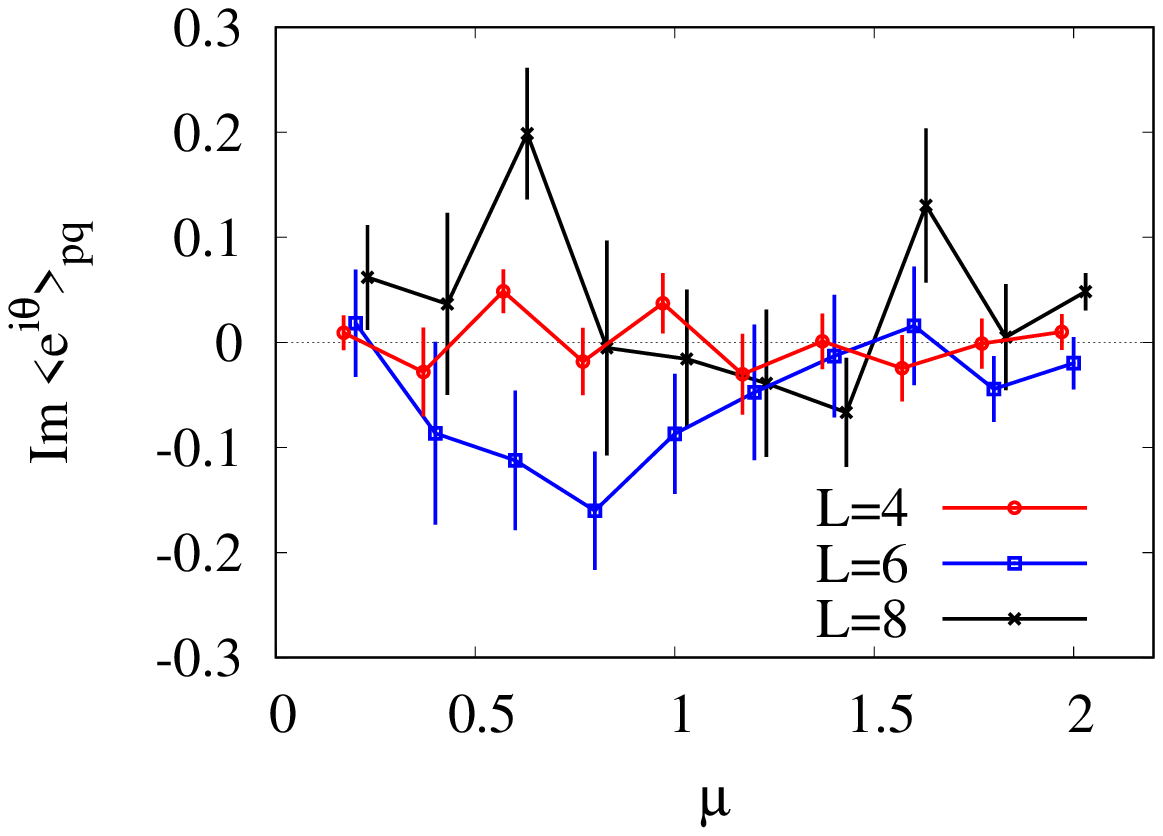}
\end{center}
\caption{
The top and middle panels show the real part of average phase factor
 without and with the optimization as a function of $\mu$.
The bottom panel shows the imaginary part of the average phase factor
 with the optimization.
Circles, squares and crosses are results for $L=4$, $6$ and $8$,
 respectively.
}
\label{Fig:apf}
\end{figure}
We first discuss the average phase factor, $\APF$.
The top (middle) panel of Fig.~\ref{Fig:apf}
shows the $\mu$-dependence of the real
part of the average phase factor, $\mathrm{Re}~\APF$, without (with) optimization.
The real part of the average phase factor without path optimization
rapidly decreases with increasing $\mu$, 
and it becomes almost zero at $\mu>1$ on a larger lattice, $L=8$.
With the path optimization, $\mathrm{Re}~\APF$ takes larger values
than that without optimization.
Particularly, results with $L=4$ show $\mathrm{Re}~\APF\sim 1$ in the range, $0 \le \mu \le 2$.
For $L=8$, $\mathrm{Re}\APF$ takes smaller values than those on the $L=4$ and $L=6$ lattices,
and takes a minimum value of $\sim 0.4$ at $\mu \sim 1.4$.
The minimum value is well above zero,
and we can safely obtain the expectation values of observables.
The bottom panel of Fig.~\ref{Fig:apf} shows the $\mu$-dependence of
$\mathrm{Im}~\APF$ with the optimization.
We find that $\mathrm{Im}~\APF$ takes smaller values than the real part.
We also note that $|\mathrm{Im}~ \langle e^{i\theta}\rangle _{pq}|$ becomes smaller
after the path optimization.  

\begin{figure}[htbp]
\begin{center}
\includegraphics[width=0.45\textwidth]{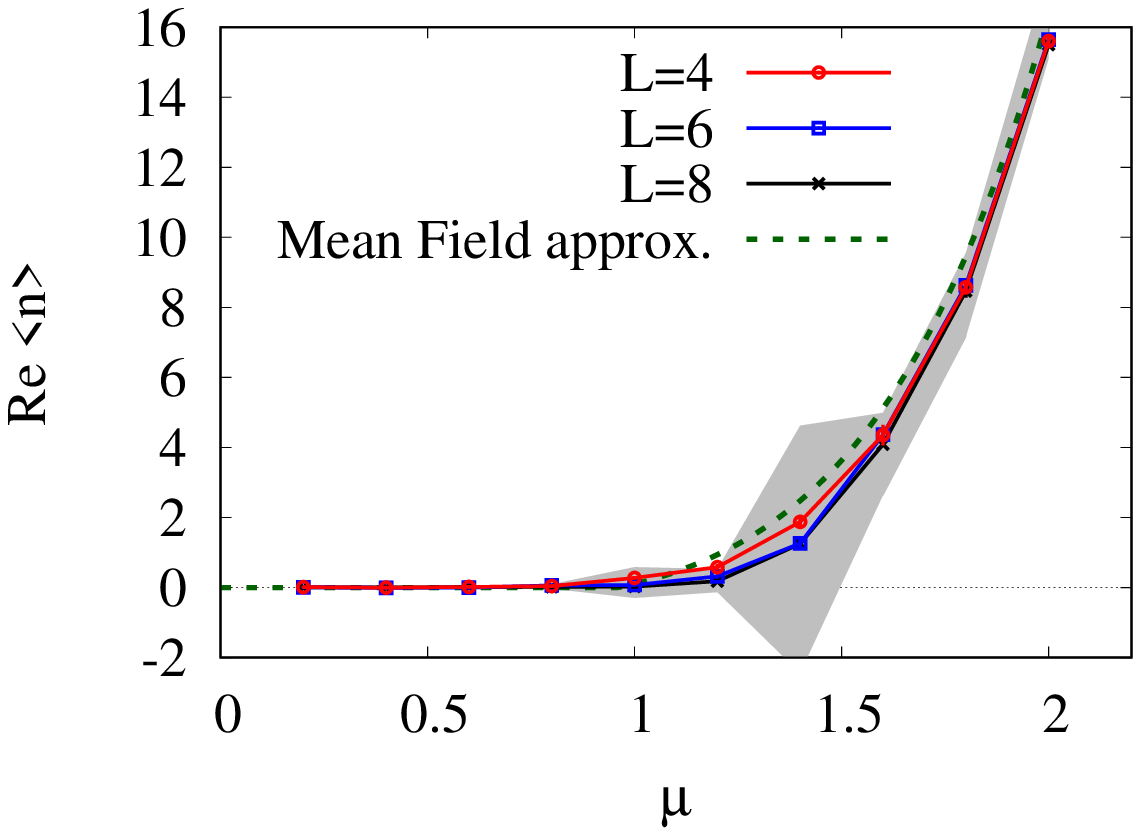}
\includegraphics[width=0.45\textwidth]{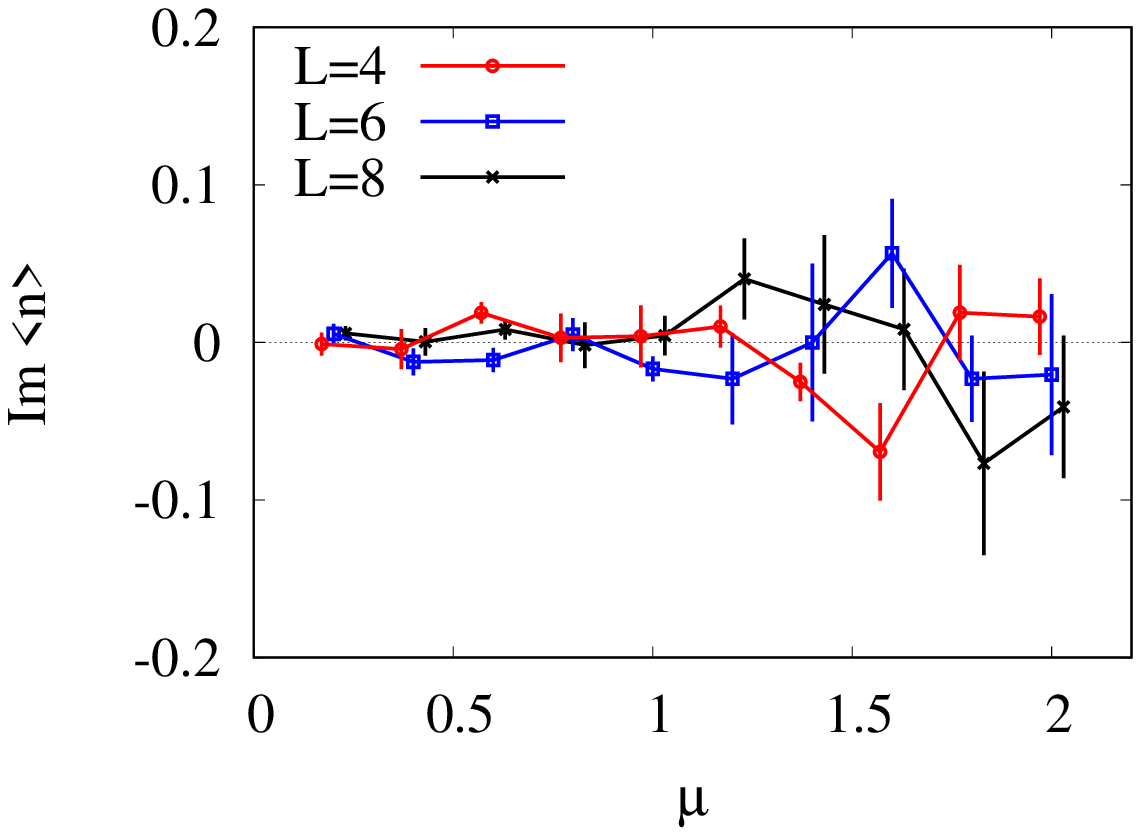}
\end{center}
\caption{
The expectation value
of the number density in the Monte-Carlo calculations
(solid lines with symbols) and in the mean field approximation (dashed line).
Shaded area shows the expectation value without path optimization method at $L=8$.
}
\label{Fig:density}
\end{figure}
Since the average phase factor is large enough with path optimization,
it becomes possible to discuss observables such as the number density.
The upper (bottom) panel of Fig.~\ref{Fig:density} shows the expectation value
of the real (imaginary) part of the number density as a function of $\mu$.
We can see that $\mathrm{Re}~\langle n \rangle$ starts to grow rapidly
around $\mu=1$.
By comparison, $\mathrm{Im}~\langle n \rangle$ is sufficiently
smaller than $\mathrm{Re}~\langle n \rangle$.
This tendency of $\mathrm{Re}~\langle n \rangle$ is consistent with the four
dimensional case; see
Refs~\cite{Aarts:2008wh,Gattringer:2012df,Cristoforetti:2013wha,Fujii:2013sra},
and references therein.
The present results are also consistent with the mean field results,
where the fields are assumed to be homogeneous and static.
In this case, the action is simplified as
\begin{align}
\frac{S}{V}=&
\left(1+\frac{m^2}{2}-\cosh\mu\right)\phi^2 + \frac{\lambda}{4}\phi^4
\ ,\\
n=&\phi^2\sinh\mu\ ,\\
\phi^2_\text{stat.}=&
\begin{cases}
0 & (|\mu| < \mu_c)\ , \\
\frac{2}{\lambda}(\cosh\mu-1-\frac{m^2}{2})  & (|\mu| \geq \mu_c)\ ,\\
\end{cases}
\end{align}
where $\phi^2=\phi_1^2+\phi_2^2$
and $\phi^2_\text{stat.}$ represents the stationary value.
The critical chemical potential is 
$\mu_c=\text{arccosh}(1+\frac{m^2}{2}) \simeq 0.962$.
The green dashed line in the left panel of Fig.~\ref{Fig:density}
shows the mean field results.
Except around $\mu\simeq1.4$, the mean field results approximately
explain the optimized path results.

\begin{figure*}[bthp]
\begin{center}
\includegraphics[width=0.24\textwidth]{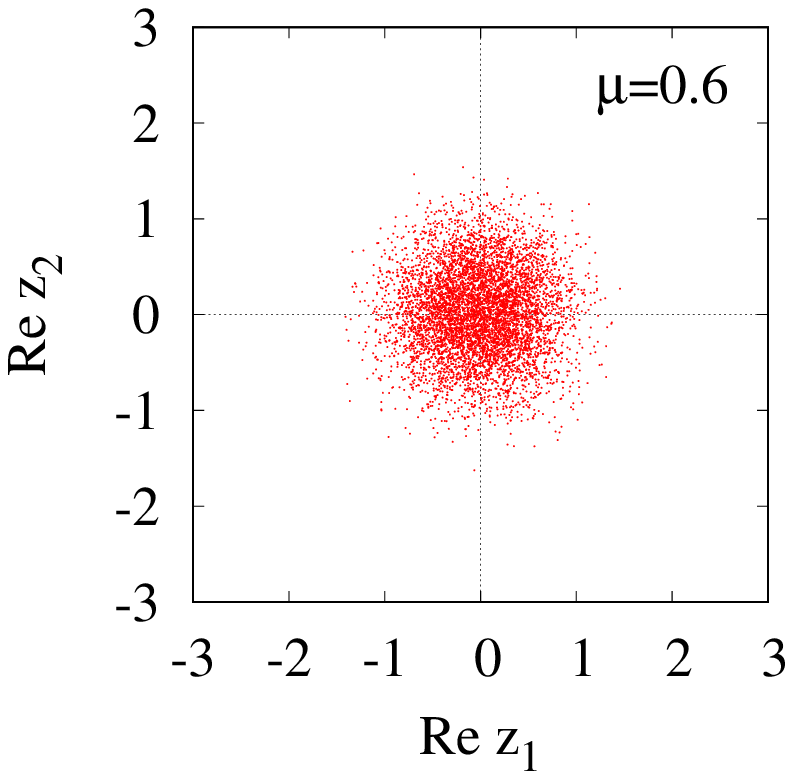}%
\includegraphics[width=0.24\textwidth]{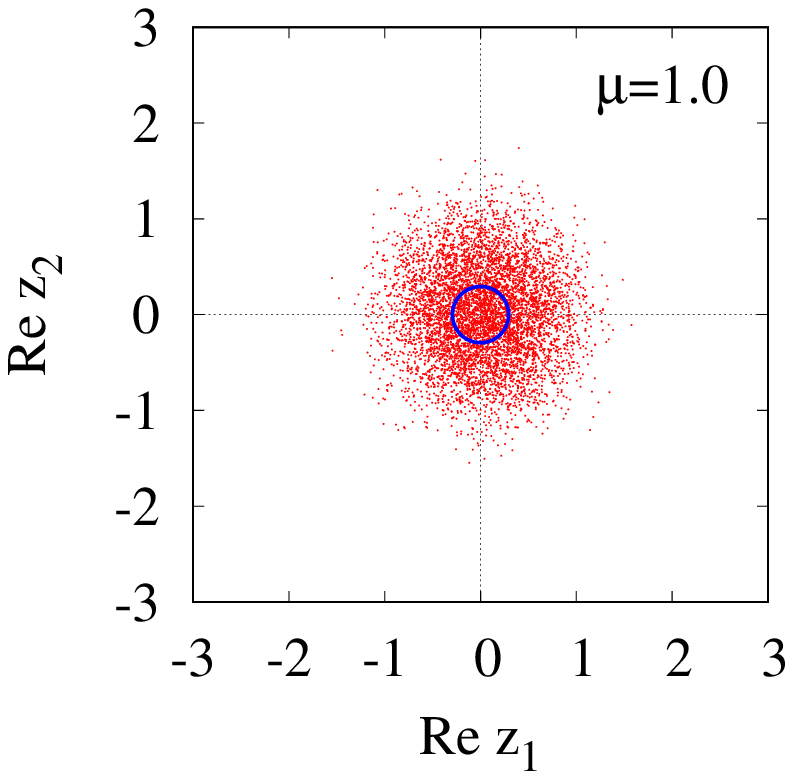}%
\includegraphics[width=0.24\textwidth]{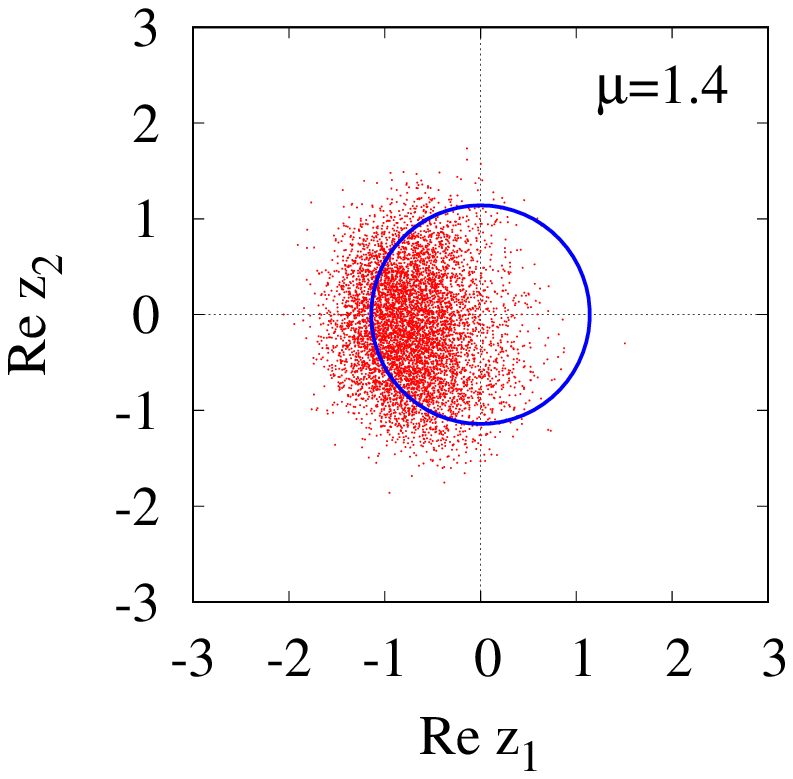}%
\includegraphics[width=0.24\textwidth]{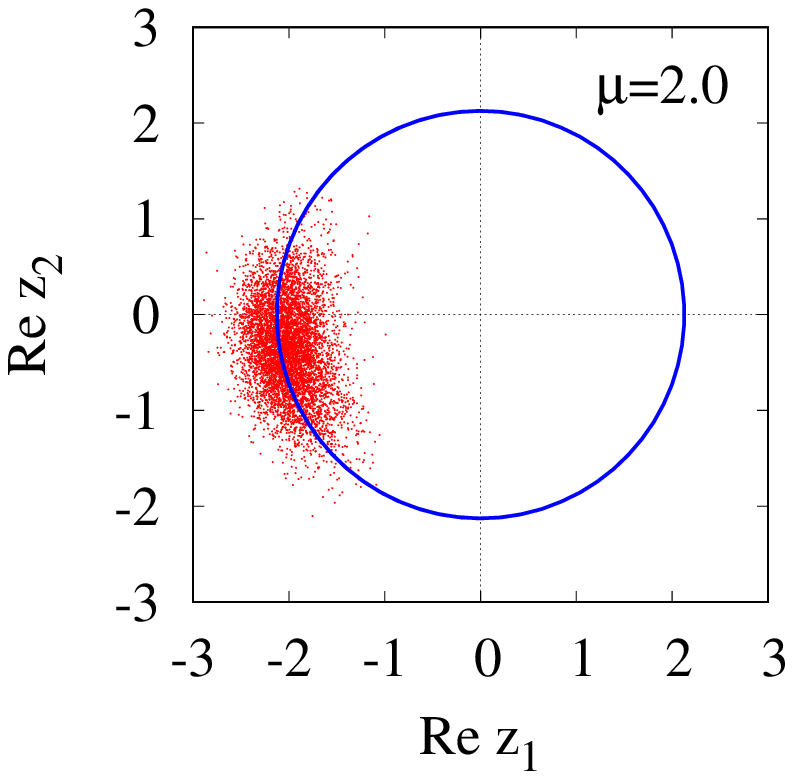}
\end{center}
\caption{
 The scatter plot of $\mathrm{Re}~z_1$ and $\mathrm{Re}~z_2$ in the case of $L=8$ at $\mu=0.6,~1.0,~1.4,~2.0$.
 Solid line shows the mean field results, i.e. homogeneous stationary points of the action.
}
\label{Fig:config}
\end{figure*}
Next, we discuss the distribution of field variables.
Figure~\ref{Fig:config} shows the scatter plot of $\mathrm{Re}~z_1$ and $\mathrm{Re}~z_2$
on the lattice sites in sampled Monte-Carlo configurations
at $\mu = 0.6, 1.0, 1.4$ and $2.0$.
In the $\mu \gtrsim 1.0$ region, we find that the distribution departs from the origin
and breaks the $\mathrm{U}(1)$ symmetry.
The deviation from rotationally symmetric distribution signals
the spontaneous symmetry breaking.
Since the field distribution is determined by the optimized path
we reach an observation that {\em the path optimization breaks the symmetry spontaneously}.

\section{Summary}
\label{Sec:Summary}

In this article, we have introduced the feedforward neural network to attack the sign problem
appearing in the quantum field theory via the path optimization method.
To obtain a better integral path in the complexified space, we have
utilized the feedforward neural network and optimized its parameters to minimize
the cost function which reflects the seriousness of the sign problem;
this procedure corresponds to increasing the average phase factor.
As an example of the quantum field theory,
we consider two-dimensional complex $\lambda \phi^4$ theory on the lattice
at finite chemical potential.

We have performed the lattice simulation on the $L^2$ lattices with
$L=4, 6$ and $8$.
We set model parameters as $m=1, \lambda=1$.
In the actual calculation, we have used ADADELTA algorithm for the optimization and
generated configurations by using the hybrid Monte-Carlo method.

It is found that the average phase factor becomes significantly larger after the optimization
compared with that for the original integral path.
In the case of $L=4$, the average phase factor is close to $1$.
By comparison, it sometimes becomes about $0.4$ in the case of $L=8$.
The average phase factor is large enough to discuss observables such as the number density.
The hybrid Monte-Carlo samples show that the spontaneous symmetry breaking can take place
at $\mu > 1.0$.
This observation suggests that the optimized path may break the symmetry spontaneously. 

Let us comment on another work on the sign problem using
machine learning developed in Ref.~\cite{Alexandru:2017czx}.
In the work, the authors have introduced the machine leaning to construct
the new integral path from the feedforward neural network which is
trained by using the few field configurations by solving the holomorphic
flow equations.
Thus, the learning is nothing but the supervised learning.
By comparison, the present path optimization method employs the unsupervised learning,
and then we do not need the teacher data and just try to
enhance the average phase factor.
It would be interesting to start from the integral path obtained with supervision,
and to further obtain unsupervised integral path with taking care
of the Jacobian phase effects.

For a future work, it is interesting to apply the path optimization
method with the artificial neural network to the lattice gauge theory
because present study means that the path optimization method can well work in the
simple quantum field theory.

\section*{Acknowledgments}
 This work
 is supported in part by the Grants-in-Aid for Scientific Research
 from JSPS (Nos. 15K05079, 15H03663, 16K05350),
 the Grants-in-Aid for Scientific
 Research on Innovative Areas from MEXT (Nos. 24105001, 24105008),
 and by the Yukawa International Program for Quark-hadron
 Sciences (YIPQS).


\bibliographystyle{ptephy}
\bibliography{ref}

\end{document}